\documentclass[preprint,prx,longbibliography]{revtex4-1}
\usepackage{subcaption}
\usepackage{graphicx}
\usepackage{comment}
\usepackage{amsmath,amssymb,amsthm} 
\usepackage{scalerel,stackengine}
\usepackage{bm}
\usepackage{verbatim}
\usepackage{float}

\begin{document}
\title{ Simulations of galaxies in an expanding Universe  with
modified Newtonian dynamics (MOND) and with modified gravitational attractions (MOGA)}
\author{S\o ren Toxvaerd}

\affiliation{ Department
 of Science and Environment, Roskilde University, Postbox 260, DK-4000 Roskilde}
\email{to be pulished in Eur. Phys. J. Plus; st@ruc.dk}

\begin{abstract}

The stability of galaxies is either explained by the existence of dark matter or caused by a modification of
	Newtonian acceleration (MOND). Here we show that the modification of the Newtonian dynamics can equally well
	be obtained by a modification of Newton's law of universal gravitational attraction (MOGA) when Newton's 
	inverse square attraction from a distant object is replaced with an inverse attraction.
        This modification 
        is often proposed in the standard model, and with the modification of the attraction caused by dark matter.
	The recently derived algorithm,   Eur. Phys. J. Plus   \textbf{ 137} 99 (2022);
	Class. Quantum Grav.  \textbf{39} 225006 (2022), for classical celestial
	dynamics is used to simulate models of the Milky Way in an
	expanding universe and with either MOND or MOGA.
	The simulations show that the galaxies with MOND dynamics are unstable whereas MOGA stabilizes the galaxies.
	The rotation velocities for objects in  galaxies with classical Newtonian dynamics
	decline inversely proportional to the square root
	of the distance $r$ to the center of the galaxy. But the rotation velocities is 
        relatively independent of $r$ for MOGA and  qualitatively in  agreement with experimentally
	determined rotation curves for galaxies in the Universe.
	The modification of the attractions  may be caused by the masses of the objects in the central part of the galaxy
	by the lensing of gravitational waves from far-away objects in the galaxy.

\end{abstract}
	
	\maketitle

\vspace{2pc}

\section{Introduction.}

The evolution of the Universe, the kinematics, and the stability of galaxies is usually explained by the presence of Dark Matter (DM) \cite{Swart2017}.
The standard model used is the $\Lambda$ Cold Dark Matter ($\Lambda$CDM)
model, in which the galaxies' rotation velocities 
 as well as their stability is ensured by  DM \cite{Bullock2017,Perivolaropoulos2022}. 
The Newtonian inverse square gravitational attraction from baryonic matter is, however, 
not included in the standard model, and this 
has given rise to a series of proposed modifications of Newton's inverse square law (ISL) for gravitational attraction
with the intention of unifying gravity with particle physics  \cite{Fischbach2001,Adelberger2003,Lee2020,Henrichs2021}.

The rotation velocities of galaxies deviate from the Classical Mechanics rotation velocities \cite{Rubin1980,Rubin1985,Gentile2011,Corbelli2000},
This behavior is taken into account in the theories that have led to the assumption of the existence of DM in the Universe  \cite{Bertone2018}.
Another attempt to explain the stability and the dynamics of galaxies is the
MOND theory, where the stability of the galaxies and their rotation velocities is caused by modified Newtonian
dynamics at small accelerations \cite{Milgrom1983}. 
The   Newtonian dynamics can either be changed by modifying the accelerations of  baryonic objects at small accelerations (MOND) or by modifying the  gravitational attractions (MOGA)
from  objects  located far away  \cite{Bekenstein1984}.
The MOND theory is reviewed in  \cite{Famaey2012}.
 
Modifications of the gravitational attraction (MOGA) have been proposed for a long time,  and modified gravity
are reviewed in \cite{Capozziello2011,Clifton2012,Mendoza2015}.
 The modification could be caused by
the general relativistic gravitational lensing (GL)
which deals with the effect of wave deflection caused by baryonic objects. GL  is reviewed in \cite{Bartelmann2010,Mukherjee2021}.
Gravitational waves were first detected in 2016  \cite{Abbott2016}.  Simultaneously detection of gravitational and electromagnetic waves
from the coalescence of binary neutron stars rules, however, out a class of MOND theories \cite{Boran2018}.

 The kinematics of the galaxies is determined by  radiation from the galaxies or by simulations.
 Simulations of galaxies  have been performed for many decades.  The dynamics of baryonic objects can be solved
 by the  Particle-Particle/Particle-Mess (PPPM)  method  \cite{Hockney1974,Klypin1983,Centrella1983}, which is a mean-field approximation where 
   each mass unit is moving in the collective field of all the others, and the  Poison equation
 for the PPPM grid is solved numerically. The dynamics are determined for  Zeldovich's adiabatic expansion of the Einstein-de Sitter  universe \cite{Zeldovich1970}.
 Later, simulations  with  large scaled computer packages with PPPM (GADGET, PHANTOM, RAMSES, AREPO)
 \cite{Springel2005,Price2018,Weinberger2020} are with  many billions of mass units.
  The evolution of galaxies has  been obtained from hydrodynamical large-scaled cosmological simulations  \cite{Schaye2010,Dubois2014,Vogelsberger2014},  
and with 
co-evolving dark matter gas  and  stellar objects \cite{Dubois2016,Ludlow2021}.
Simulations of galaxies with MOND dynamics have also been performed by  (approximated)  Poisson solvers and  \cite{Angus2014,Angus2014a,Lughausen2014}.
Cosmological simulations of galaxy formation are reviewed in \cite{Vogelsberger2020}.

Here we simulate models of galaxies with the modified Newtonian dynamics, MOND,
and with modified gravitational attraction, MOGA. The MOND acceleration
is obtained from the interpolation function(s)  \cite{Milgrom1983,Gentile2011}, and the modified gravitational
attraction, MOGA, is obtained by replacing Newton's gravitational inverse square attraction (ISL) with
an inverse attraction (IA) for large distances. The simulations of galaxies in an expanding Universe are performed by use of a recent extension of Newton's discrete algorithm
\cite{Newton1687,Toxvaerd2020,Toxvaerd2022,Toxvaerd2022a}. The algorithm is absolutely stable and \textit{without any approximations}, and the dynamics with the discrete algorithm have
the same invariances as Newton's analytic dynamics, and thus the discrete dynamics is exact in the same sense as the exact solution of Newton's   analytic second-order
differential equations for classical
celestial mechanics. The  dynamics
with Newton's discrete algorithm is reviewed in \cite{Toxvaerd2023}, and the algorithm and the proof that Newton's discrete dynamics has the same invariances as his analytic
dynamics is given in the Appendix. The advantage of using Newton's discrete algorithm is that the classical dynamics are without any approximations, and the disadvantage is
that it is only possible to perform the exact long-time simulations for small ensembles of objects in bound rotations around their common mass center.

The Newtonian dynamics with MOND and with  MOGA are formulated in the next section. The acceleration in MOND is modified when
the acceleration of an object is below a certain threshold, $a_0$, and $independent$ of the individual attractions from the objects
which together cause the small acceleration. This modification breaks Newton's third law and unlike MOGA, MOND
 no longer conserves the momentum and angular momentum for a conservative system(see the Appendix, Eq. (A.7), Figure A.1, and \cite{Felten1984}), and the present simulations
reveal that  MOND is unstable. Section 3 presents the results of the
MOND and MOGA simulations of galaxies for times corresponding to the age of the Universe.
The simulations show that the MOND dynamics is unstable and releases the bound objects in the galaxies with time, whereas MOGA not only stabilizes the bound rotations
of the objects in the galaxies but it also changes the rotation velocities and is qualitatively in  agreement with experimentally
	determined rotation curves for galaxies in the Universe.

\section{Modified acceleration, MOND, and modified gravitational attraction, MOGA.}

Newtonian dynamics for celestial objects is given by Newton's classical dynamics  
\begin{equation}
	\textbf{F}_{i}(t)=m_i \textrm{a}_i \hat{\textbf{a}}_i(t) \\
\end{equation} 
and his ISL law  of  gravitation
\begin{equation}
	\textbf{F}_{i}(t)=\textrm{F}_i \hat{\textbf{a}}_i=  -\sum_{j \ne i}^N {\frac{m_i m_j G}{r_{ij}^2(t)} \hat{\textbf{r}}_{ij}(t)}
\end{equation}
for the force $\textbf{F}_{i}(t)$ and the  acceleration $ \textbf{a}_i(t)= \textrm{a}_i \hat{\textbf{a}}_i(t)$ for the object $i$ in the ensemble of $N$ objects,
caused by  baryonic objects $j$ with mass $m_j$ at  distances $r_{ij}(t)$  and at time $t$.

The first equation is Newton's second  law for the relation between a force  acting on an object, and Newton postulated Eq. (1)
in the very first part of $Principia$ and continued by derived  Eq. (1) at page 37  from a discrete analog formulation of the change in the position
by a force impulse $\textbf{F}_{i}(t)$ at time $t$ (see Apendix A).
Much later   Newton starts on page 401 in $Principia$ by formulating four principles for philosofical rules in natural science.  Rule I in an
 English translation reads:\\
\textit{ We are to admit no more causes of natural things than such as are both true and sufficient to explain their appearances.}
After the formulation of the philosophical principles, he succeeds in formulating his law of universal gravitation, given by Eq. (2).
The law was obtained from experimental data for the positions of the planets in the Solar system and the positions of the Moon. 
MOND and MOGA acknowledge the philosophical principles in an attempt to explain the stability and behavior of the dynamics in the Universe.
MOND by modifying the acceleration and MOGA by modifying the ISL.

The  MOND theory was published in 1983 by M. Milgrom \cite{Milgrom1983}.
The acceleration $\textbf{a}_i(t)$ in classical Newtonian dynamics is modified for a small acceleration
caused by the sum of interactions with the $N-1$ baryonic objects located at large distances $r_{ij}$ from $i$. The transition
from the Newtonian acceleration to MOND occurs for a small acceleration $a_i \approx a_0 =\mid\textbf{a}_0\mid$. This can only happen if
 $all$ the attractions  with object $i$ from  the other $N-1$ baryonic objects are for  large distances so that the sum of attractions results in a
 small acceleration $ a_i< a_0$
\begin{equation}
	a_i \rightarrow \mu(a_i/a_0)a_i. 
\end{equation}
 According to Bekenstein and Milgrom \cite{Bekenstein1984} the modification can either be performed by modifying the  law of inertia (MOGA) or  by modifying the acceleration (MOND).
The modification, $\mu $,  of the acceleration  $\mid\textbf{a}_i\mid$
is obtained  with the ''standard interpolation function" 
\begin{equation}
	\mu(\textrm{a}/a_0)=\sqrt{ \frac{1}{1+(\frac{a_0}{\textrm{a}})^2}},
\end{equation}
The MOND acceleration is given by
\begin{equation}
	\textrm{F}_i/m_i= \textrm{a}_i \sqrt{ \frac{\textrm{a}_i^2}{\textrm{a}_i^2+a_0^2}},
\end{equation}
or the interpolation function  proposed by \cite{Gentile2011}
\begin{equation}
	\mu(\textrm{a}/a_0)=\frac{\mid\textrm{a}\mid}{ \mid\textrm{a}\mid+a_0},
\end{equation}
with the modification given by
\begin{equation}
	\textrm{F}_i/m_i	=
	\textrm{a}_i \frac{\mid\textrm{a}_i\mid}{ \mid\textrm{a}_i \mid+a_0},
\end{equation}
and acceleration 
\begin{equation}
	\textrm{a}_i(\textrm{MOND})  =\frac{ \textrm{F}_i}{2m_i}(1+\sqrt{1+4 m_i a_0/\mid\textrm{F}_i\mid}).
\end{equation}	

The MOND modifications, Eq. (5) or Eq. (7), change the acceleration
  from the classical Newtonian  acceleration $\mid\textbf{a}\mid  >> a_0 $ at short distances to  a modified acceleration
\begin{equation}
	\mid\textbf{a}_i(\textrm{MOND})\mid= \sqrt{ \mid\textrm{F}_i\mid a_0/m_i}
\end{equation}
for $ \mid\textbf{a} \mid << a_0$.

The asymptotic modified acceleration for an isolated object $i$  
 with only one gravitational interaction, $\textrm{F}_i(r_{ij})=-m_im_jG/r_{ij}^{2}$, with another object, No. $j$, is obtained from F$_i(r_{ij})$ and Eq.(9) as
\begin{equation}
	\textrm{a}_i(\textrm{MOND})= - \frac{ \sqrt{m_jG a_0}}{r_{ij}}.
\end{equation}
MOND is a modification of  Newtonian acceleration. But in this case, the modification might as well  be formulated as a modification
of Newton's ISL law of universal gravitational attraction, where the inverse square attraction asymptotically is replaced with an inverse attraction (IA).
If this modified gravitational attraction is a  universal law, MOGA, the gravitational force is modified to
 
\begin{equation}
	\textbf{F}_i(\textrm{MOGA})=-\sum^N_{j \ne i}\frac{m_i m_j G}{r_{ij}^2}(1+    \frac{r_{ij}}{r_0}) \hat{\textbf{r}}_{ij}(t)
	\end{equation}
with $r_0=\sqrt{m_jG/a_0}$.

The modified gravitational attraction, Eq. (11), is  a specific example of a general modification of the force field \cite{Capozziello2011,Bekenstein1984}. 
Newtonian dynamics and Newton's discrete algorithm  used in the next section are time reversible,
symplectic and with the dynamical invariances:  momentum, angular momentum, and energy for a conservative system \cite{Toxvaerd2023}. 
MOGA   with Newton's discrete algorithm, maintains these qualities,
whereas MOND does not conserve momentum and angular momentum (see Appendix A.2, Figure A1 and \cite{Felten1984}).
The Hubble expansion of the space destroys the invariances \cite{Toxvaerd2022a}, but 
 MOND and MOGA dynamics are, however, still time reversible. 

 Modifications of the Newtonian ISL attraction have been proposed for a long time 
 in an attempt to obtain the stability of galaxies
  by the standard model \cite{Fischbach2001,Adelberger2003,Henrichs2021,Capozziello2017,Finch2018},
   and the dynamics of local group of galaxies (Milky Way and Andromeda) have been described by modified   Newtonian  attraction  \cite{Benistry2023,Benistry2023a}.
  For a review of extended theories of gravity see \cite{Capozziello2011,Capozziello2024}.

 The modified Newtonian gravitational potential $u(r)$ is often  modified by a Yukawa potential \cite{Cardone2011,Chen2016,Baeza-Ballesteros2022,Benistry2023a}
 \begin{equation}
	 v_{ij}(r)=u(r_{ij})[1+\alpha exp(-r_{ij}/\lambda)],
 \end{equation}
and the corresponding modified gravitational forces are
 \begin{equation}
	 \textrm{F}_i = -\sum^N_{j \ne i}\frac{m_i m_j G}{r_{ij}^2}[1+\alpha exp(-r_{ij}/\lambda)(1+r_{ij}/\lambda)]. 
 \end{equation}
 The parameter $\lambda$ corresponds to the parameter $r_0$ in MOGA, and 
   a possible deviation from the ISL gravity was investigated for $\lambda$ in
   the range $\lambda \in [30,8000]$ nm  \cite{Chen2016,Bimonte2021,Baeza-Ballesteros2022}, and the rotation velocities of stars in the Milky Way
   were used to determine a possible  Yukawa correction, Eq. (12) to the gravitational attraction \cite{Henrichs2021}. So far, however, there is no
   direct experimental evidence for deviation from the ISL gravitational attraction. 
\begin{figure} 	
	\begin{center}	
\includegraphics[width=8cm,angle=-90]{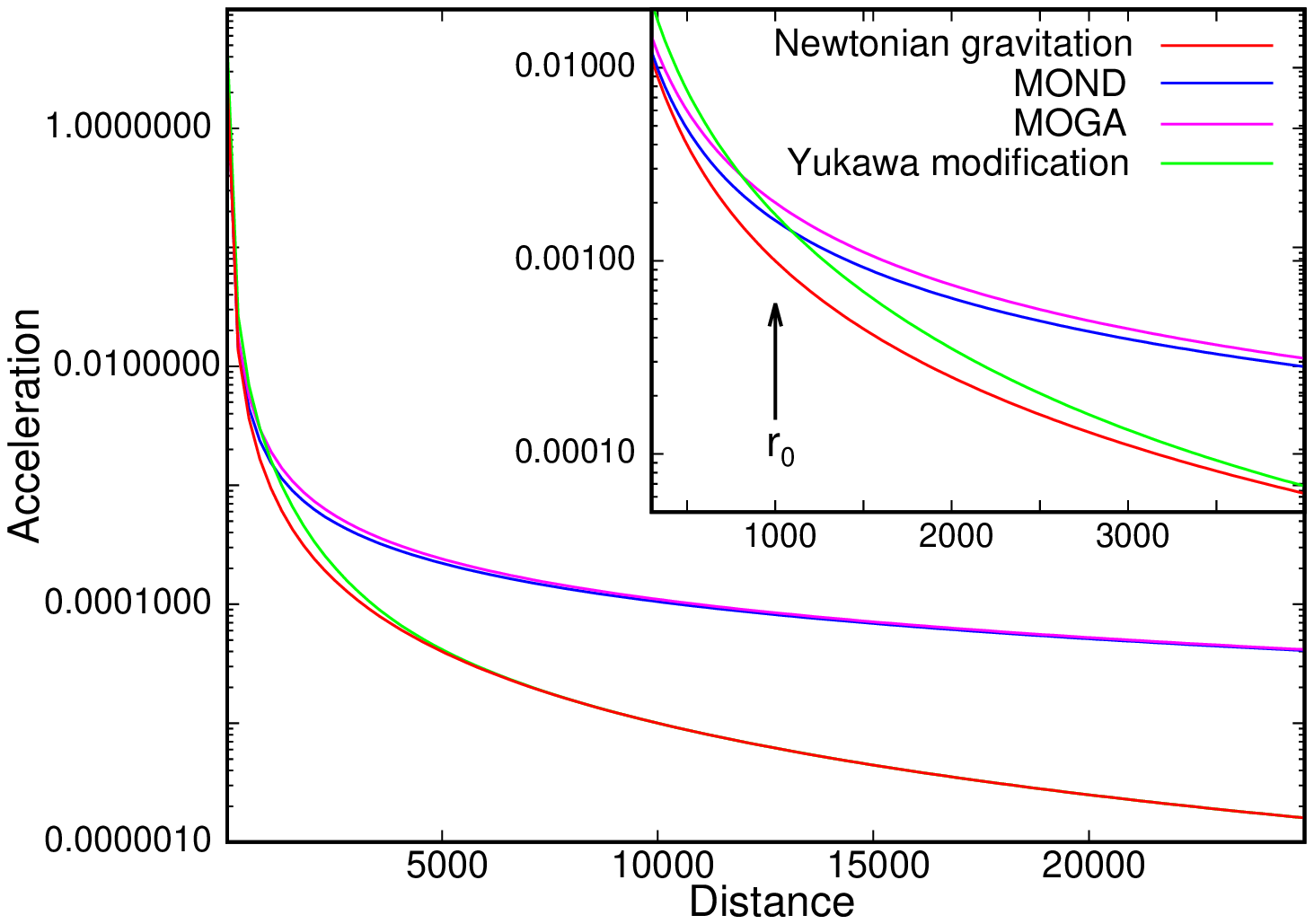}
		\caption{ An illustrative figure of the modified  accelerations a=F/$m$ of an object
		as a function of the distance  $r$ to a heavy object with a mass
		 $M=1000m$. Length unit is in parsec and force (F=$mMG/r^2)$ is in units  of $m$ and $G$.
		 The MOND acceleration (Eq. 8) is in blue, the  MOGA acceleration (Eq. 11) is in magenta, and the
 acceleration with the  modified Yukawa attraction (Eq. 13) is in  green. The accelerations  are for $r_0=\lambda$=1000 parsec.
The Newtonian acceleration with the ISL (Eq. 2) is shown in red. The inset enlarges
the differences at $r \approx r_0$.	}
\end{center}
\end{figure} 
\begin{figure} 	
	\begin{center}	
\includegraphics[width=8cm,angle=-90]{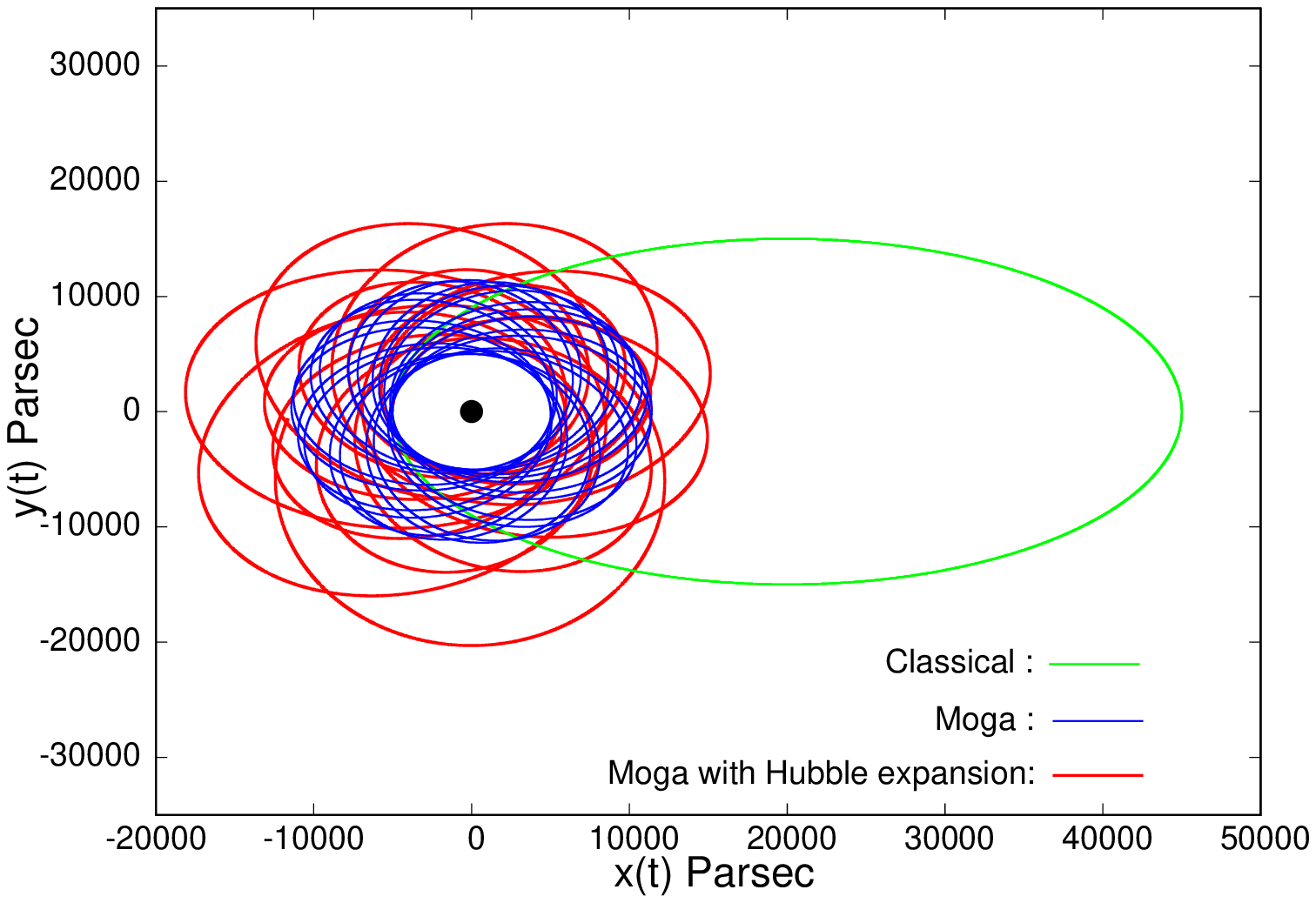}
		\caption{ The orbits of an object near a mass center in black  which is thousand times heavier than the object.
		The dynamics was started at Perihelion at a distance to the center  $r($Perihelion)=5000 pc
		and with an excentricity $\epsilon=0.80$. The green ellipse is the classical
		orbit. In blue are the first  38 corresponding (revolving) orbits with  MOGA and with $r_0=25000 $ pc.
		The corresponding 26 revolving orbits in red are for the MOGA with the Hubble expansion included in the dynamics \cite{Toxvaerd2022a}.}
\end{center}
\end{figure} 

An illustrative example of 
the different accelerations are shown in Figure 1. (Units are with $m_i=G=1$ and length unit $1$ corresponds to 1 pc for a galaxy.
For units see \cite{Toxvaerd2022a}). The accelerations are for $one$ object attracted by a heavy mass center with
a mass that is a thousand times heavier  than the object and for a modification
distance $r_0=\lambda=1000$, which corresponds to 1 kpc for the model for the Milky Way in the next section.
The Yukawa modification is shown in green, the MOND acceleration is in blue, and the MOGA acceleration in magenta. The MOND and MOGA modifications
are rather similar in this case, but MOGA and MOND for galaxies with many gravitational objects
are, however, very different.
The difference between MOGA and MOND for the dynamics of a galaxy originates from the summation of interactions between the objects. 
The  acceleration in MOND  is modified if the interactions of an object with $all$ the other baryonic objects result in an
acceleration below a certain threshold, whereas the contribution to the acceleration   
from the attractions of all faraway objects is modified in MOGA. The green curve is the force from the modification with the
Yukawa potential, Eq. (13), and with $\alpha=1$ and  $\lambda=$1000.
If there exist dark matter this modified force will be effective on a much shorter length scale 
with baryonic attractions from dark matter in the galaxies and their halos.

The mean rotation velocities of stars in galaxies are rather constant and independent of the distance to the
centers of rotation as opposed to a system of rotating baryonic objects with classical Newtonian dynamics \cite{Milgrom1983,Gentile2011}.
This led Milgrom to propose the modification, Eq. (4) (or Eq. (6)) of the Newtonian classical acceleration,
and with the asymptotic modification Eq. (9). By adjusting the modified  acceleration to one (isolated) baryonic object in
rotation at a gravitational center he determined a value for the constant $a_0$.
Milgrom
found $a_0 \approx 1.2 \times 10^{-10}$ ms$^{-2}$ to be optimal, and later investigation of the
rotation curves for stars in 12 galaxies confirmed this value  \cite{Gentile2011}. 
For a MOND modification caused by only one baryonic object  with mass $m_j$ equal to the mass of
 our Sun: $m_j=m_{\textrm{Sun}}$=1.989$\times 10^{30}$ kg and with the gravitational constant $G=6.647 \times 10^{-11}$ N m$^{2}$ 
 kg$^{-2}$ the constant $a_0$ corresponds to a distance 
 \begin{eqnarray}
	 r_0=\sqrt{\frac{m_jG}{a_0}}= \sqrt{\frac{1.99 \times 10^{30} \times 6.65 \times 10^{-11}}{ 1.2 × 10^{-10}}} \nonumber \\
	 = 1.05 \times 10^{15} \textrm{m}=\frac{1.05 \times 10^{15} \textrm{m}} {3.09 \times 10^{16}  \textrm{m} \textrm{parsec}^{-1}} \nonumber \\
	 =0.033  \textrm{parsec},
 \end{eqnarray}	 
The example is for one interaction from an object with mass equal to our Sun. The modification distance is proportional
with the square root of the masses, $m_j$, and the distance is much larger for a modification caused by many heavy objects far away. 
The Milky Way is a barred spiral galaxy. The extension of the barred disk is 
2.5-3 kpc \cite{Rix2013}  and the extension of the halos is  $\approx$ 100-300 
kpc \cite{Deason2020,Li2021}, so the galaxy will be affected by a modification of the gravitational attractions or the accelerations. 

 MOND and MOGA dynamics are equal in the case of only one object  attracted by a heavy mass center according 
 to Eq. 11. Figure 2 shows the dynamics for one object in  rotation around a heavy mass center with a mass one thousand
 times bigger than the object's mass. The green curve is the elliptical orbit of the object
 with pure classical Newtonian dynamics. The dynamics was started in Perihelion with a distance to the
 mass center $r($Perihelion)=5000, which corresponds to 5000 Parsec for the Milky Way, and with an excentricity 0.8
 by which the object's distance to the center is 45000 in Aphelion.
 (The relations  between  the units for length  $l$:  1 pc, time $t$: 1 Gyr   and Hubble expansion coefficient
$H=72.1 \pm  2.0 \  \textrm{km}  \textrm{s}^{-1} \textrm{Mpc}^{-1}$ \cite{Soltis2021}	 in the Universe and  the corresponding
units for the MD models of galaxies are:
 1 pc $\hat{=}$ 1 MD length unit, 1 Gyr $\hat{=} 6\times 10^5$  MD  time units and 
$ 72.1 \textrm{km}  \textrm{s}^{-1} \textrm{Mpc}^{-1} \hat{=} 5.\times 10^{-8}$ in MD units. For determination of these relations see \cite{Toxvaerd2022a}.)
 The blue curves are the first 38 revolving orbits with MOND=MOGA and with the modification distance $r_{0}=25000$ in Eq. (11).
 The modified dynamics results in stable revolving orbits, but with a smaller mean distance
 to the mass center \cite{Newton1687,Toxvaerd2022}.
 The red curve is the first 26 corresponding orbits for MOGA dynamics with a Hubble expansion equal
 to the Hubble expansion of the Universe \cite{Toxvaerd2022a}. The dynamics
 still exhibit revolving orbits,  but with an increasing mean distance.
 MOGA with Hubble expansion performed 64 regular orbits before the expansion
 released the object from the mass center, and
 the object performs many more orbits for stronger mass centers  before its release.
 The Milky Way has performed  $\approx$  60 rotations
 after its creation more than thirteen billion years ago and the Hubble expansion does not at all affect the
 stability of the revolving orbits for MOGA  with heavier mass centers in times corresponding to the age of the Universe.
 The total mass of the Milky Way is extimated to $2.06 \times 10^{11}-5.4 \times 10^{11}$ in unit of the mass of the Sun \cite{Jiao2023}. 

\section{Simulations of galaxies with MOGA and with MOND.}

Simulations of galaxies have been performed for many decades  \cite{Aerseth1963},
but the simulations here of models of galaxies deviate from the main part of the simulations in that they are pure Newtonian N-body simulations.
 Such simulations have, however, also been simulated for a long time. But
 one has performed a series of approximations,  such as variable time steps and mean field approximations, because these simulations are very time-consuming.
 But the approximations ruin the exactness of the simulations. The
 simulations below are exact N-body Newtonian simulations without any approximations. The algorithm with the conserved dynamic invariances 
 is given in the Appendix and in a recent review article about
  Newton's exact discrete dynamics \cite{Toxvaerd2023}.

A galaxy and the Milky Way contain hundreds of billions of stars, and a substantial amount of baryonic 
gas \cite{Gupta2012,Bergman2018,Jiao2023}, and
it is not possible   to obtain the  exact  dynamics with MOGA or with MOND of a galaxy
with this number of objects.  We have instead of simulated
  models of small ''galaxies" of hundred of objects in
   orbits around their center of gravity, and in an expanding space with various values of $r_0$ or $a_0$.
A recent article describes how a system of baryonic objects with Newtonian discrete dynamics spontaneously creates a system with
 the objects in rotation about their center of gravity \cite{Toxvaerd2022}. The algorithm is extended to also include the dynamics with the Hubble expansion of the space
\cite{Toxvaerd2022a}.

The  algorithm is used to simulate  models of a galaxy with the Hubble expansion and
with MOGA and MOND, respectively.
An ensemble of gravitational objects with Newtonian dynamics 
at  time $t$=0  might spontaneously create a ''galaxy" system with many of the objects
in a bound rotation around the center of gravity, and
one can either start the simulations with MOGA or MOND at $t=0$, or alternatively  at a later time $t$  
where a Newtonian galaxy is created and it is in a rather stable state. 
The data reported below for  twelve galaxies are started from a stable Newtonian galaxy. The twelve galaxies with the data reported below   are
for $r_0= 1, 10, 100, 1000, 10000, 100000$ and $a_0 = 1, 10^{-2}, 10^{-4}, 10^{-6}, 10^{-8}, 10^{-10},$  respectively, and each (stable) galaxy are simulated  $3.2 \times 10^9$ time steps corresponding
to $t=8 \times 10^6 \approx$ 13.4 Gyr or the age of the Universe. Each simulation with the  $3.2 \times 10^9$ time steps 
 took $\approx$ 1000 hours on a fast CPU in the CPU-cluster, and the total amount of simulations for different values of $r_0, a_0, H$, and start configurations is fifty-three. 

\begin{figure} 	
\begin{center}	
\includegraphics[width=8cm,angle=-90]{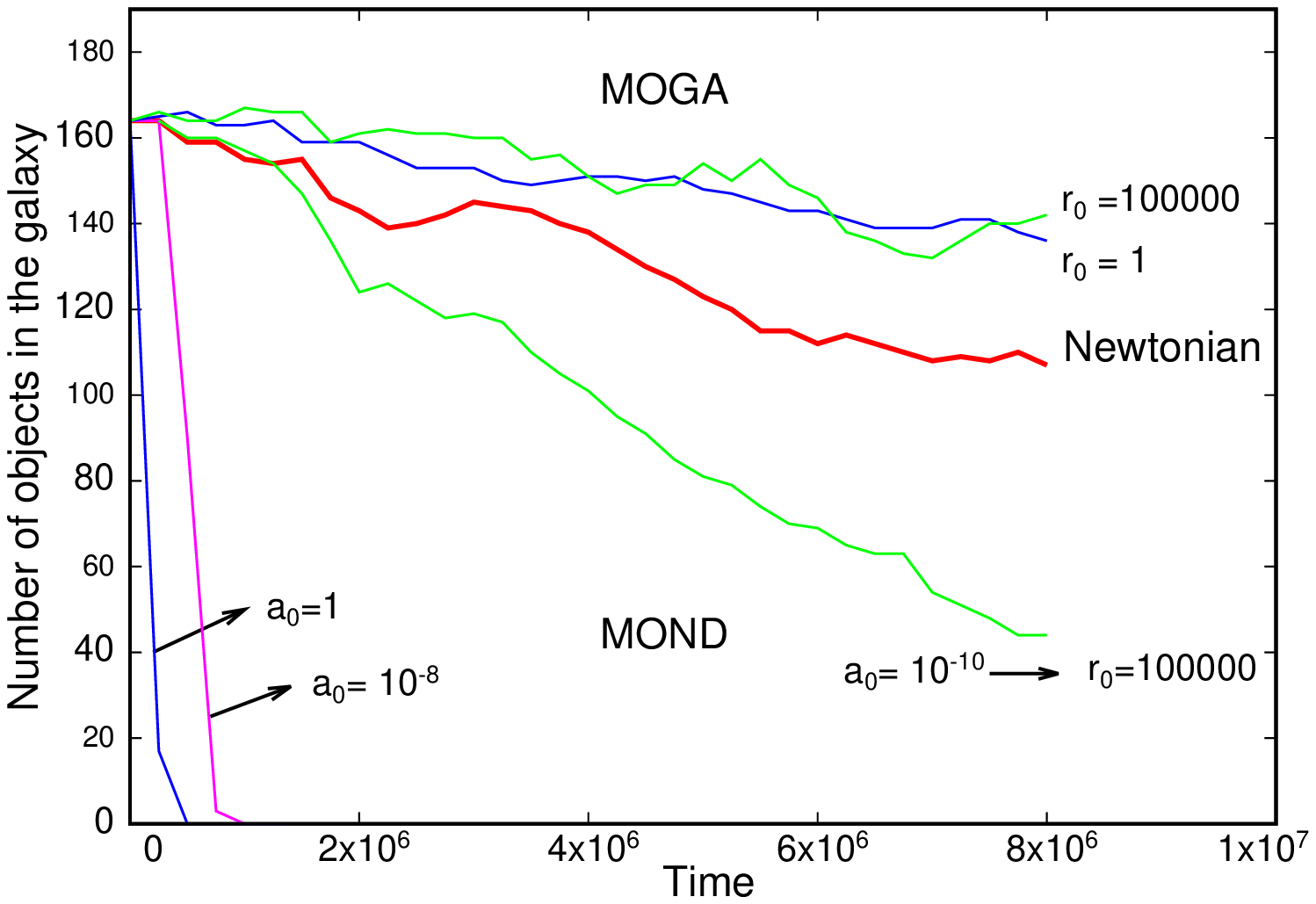}
	\caption{The number of  objects  with mean distances $r(\textrm{mean}) < 100000$ parsec to the center of mass  as a
	function of time  $t \in [0,8 \times 10^6] \approx$ 13.4 Gyr 
	where the  galaxies are exposed to either MOND or MOGA. 
The red curve is for Newtonian dynamics without any modification of accelerations or attractions. The green
curves are for a very weak/long range modification $ a_0= 10^{-10}$ (MOND) corresponding to $r_0=100000$ (MOGA), and the blue
curves are for a strong/short range modification $ a_0=r_0=1$. The magenta curve is MOND dynamics with $ a_0= 10^{-8}$.}
\end{center}
\end{figure} 

\subsection{ Stability of   galaxies with MOGA or with MOND.}
The dynamical effect of the modifications of the accelerations or gravitational forces is obtained by
exposing the objects in a  galaxy to MOGA or to MOND. The results reported below are obtained by exposing the objects in a  Newtonian galaxy
which is in a rather stable state and with an occasional release of an object before the modifications (Curve in red in Figure 2).
At the start of the modifications the system  contains 460 objects with 165 with
a mean distances $r(\textrm{mean}) < 100000)$ parsec to the center of mass, 
and 295 objects with mean
distances $r\textrm{(mean}) > 100 000 \approx 100$ kpc.
The number of  objects with mean distances $r(\textrm{mean}) < 100000)$  in the galaxy after the change of dynamics is
shown in Figure 3.

 The stars in the halos of a galaxy are not in a stable and bound rotation, and this fact has led to the hypothesis
about the existence of dark matter in the Universe. The simulations of Newtonian galaxies
show, however, that the release of stars in the outer edge of a  Newtonian galaxy is
rare, and that the  Newtonian galaxies are rather stable over time periods of many Gyr (red curve in Figure 3) \cite{Toxvaerd2022a}.
By changing the Newtonian dynamics to MOND the enhanced MOND acceleration destabilizes 
the galaxy and results in a release of the bound objects, whereas the MOGA dynamics has the opposite effect.
The number of bound objects  with $r_0=a_0$=1
is shown in blue, and the number with $a_0=10^{-10}$ (MOND) and correspondingly  $r_0$=100000 (MOGA) is shown in green in Figure 3.
MOGA stabilizes the
galaxy even for a very weak modification with $r_0 \approx 10000- 100 000$  corresponding
to a modification of the attractions at distances $\approx$ 10-100 kpc for the Milky Way. The galaxies with MOGA still release objects, but this 
is very rare and the galaxies contain still more than 135 bound objects after a time $\Delta t= 8 \times 10^6 $  corresponding to 13.4 Gyr, or the age 
of the Universe.  The simulations are performed for $r_0= 1 (136), 10 (143), 100 (142), 1000 (156), 10000 (156)$ and  $100000 (142)$, respectively,
and with the number of  objects with mean distances $r(\textrm{mean}) < 100 000$ at the end of the simulations
indicated in the parentheses. The number of objects in the Newtonian galaxy with $r_0 =\infty$ at the end of the simulation
(red curve in Figure 3) is $N=$107, so the MOGA galaxies contain significantly more
bound objects than the Newtonian galaxy. 
The stability of the galaxy given by the number of bound objects with MOGA is not sensitive to the range $r_0$ of the modification of the gravitational attraction.

The number of bound objects ($r(\textrm{mean}) < 100000$) with MOND dynamics is also shown in Figure 3.
The MOND dynamics releases the objects even for a very low threshold for the modified acceleration, given by $a_0=1/r_0^2$.
It was only possible to maintain some bound objects for a very weak modification of the Newtonian
accelerations with $a_0=10^{-10} \rightarrow r_0=100000$
(green MOND curve in Figure 3).
All the bound objects were released for a stronger  MOND modification of the acceleration.

 Galaxies with  MOGA or with MOND dynamics were simulated with other start distributions of the baryonic objects and for $H=0$ (i.e. without a Hubble expansion of the space).
 The simulations showed unanimously, that 
 MOGA dynamics has  a stabilizing effect on the   objects in the  galaxies even for a big value of $r_0$ corresponding to that only the
  stars in the halos of a galaxy are affected by the modified gravitational attraction.
All the MOND simulations were unstable and released the bound objects sooner or later. MOND does not conserve the angular momentum of the ensemble (see Appendix A.2.,
Eqn. (A.7) and (A.9) and Figure A1), and the increased MOND acceleration has the opposite effect, it  destabilizes the bound objects in the
 galaxies even for the  very small value of $a_0=10^{-10}$ and the  MOND galaxies release the objects from their regular orbits in the galaxies.
 The same was thru for dynamics  without a Hubble expansion ($H=0$), where the galaxies were stable for MOGA, but unstable for MOND.
 
\subsection{ Rotation velocity of the stars in  a galaxy.}
There is a   discrepancy between  the mean rotation velocities of the stars   in the galaxies  and the corresponding 
mean velocities of the objects in the simulated galaxies with pure  Newtonian dynamics.
 The mean rotation velocities of the stars in a galaxy as a function of the distance to the center of rotation increase for short distances, but  are rather constant
 at larger distances to the center of the galaxies.
 \cite{Corbelli2000,Gentile2011,Famaey2005}.
But the  velocity $v(r)$ of objects with pure Newtonian dynamics
in mean declines as $v(r) \approx (M/r)^{0.5}$ with the distance $r$ to the mass center  of the galaxy with mass $M$. The rotation velocity of 
stars in the Milky Way at the distances $r \in [6.72, 8.40]$ kpc is $v(r) \in [203, 240]$  km s$^{-1}$ \cite{Camarillo2018}. One of the results 
in \cite{Toxvaerd2022a} from the simulation of galaxies with pure Newtonian dynamics was,
that the velocities of objects for distances $r \in [15000,100 000]$ to the center were not located near a  line
with  $v(r) = (M/r)^{0.5}$, but were diffusely distributed.
But a determination of the rotation velocities over  a longer time interval
  with the  root mean square (rms), obtained from 22 consecutive time intervals of
the mean velocities,  reveals
that the  distribution of the mean rotation velocities  is rather Newtonian,
and disagrees with the $\approx$ constant rotation velocity in galaxies (Figure 4).

\begin{figure} 	
	\begin{center}	
\includegraphics[width=8cm,angle=-90]{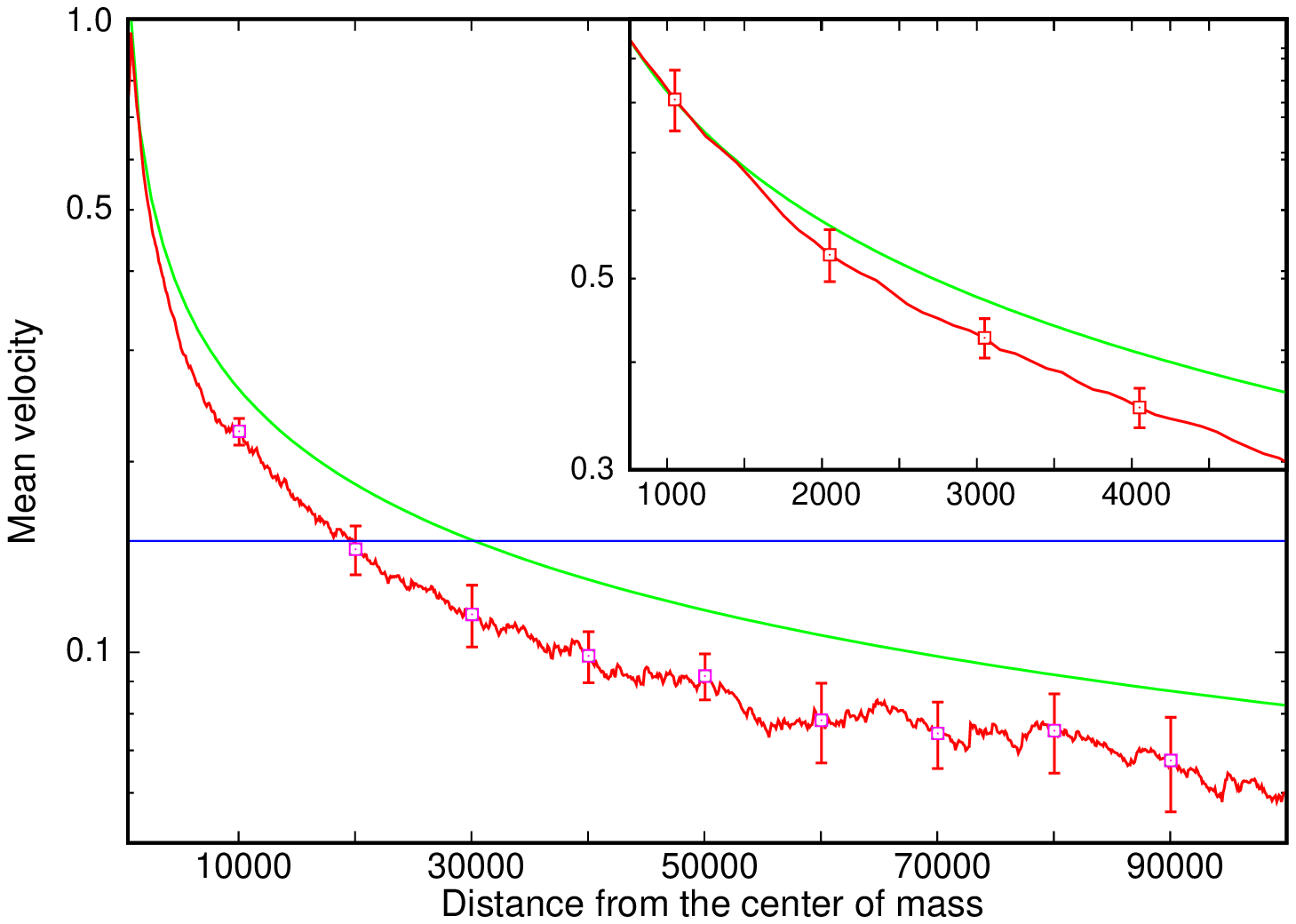}
	\caption{ Mean rotation velocities of the objects in the galaxy with Newtonian dynamics in red (i.e. without a modification of accelerations or attractions).
		Distances are in parsec and velocities are in MD units
		where $v(r)=0.15$ corresponds to $v=220$ km/s in the Milky Way.
		In green is
	the Newtonian rotation velocity
		$v(r)=(M/r)^{0.5}$ for one bound object attracted by a heavy mass center. The blue line is $v(r)=0.15$.
		The standard deviations for (representative) selected distances are obtained from 22 consecutive time subsets of
the mean rotation velocities in the subsets. The inset shows the mean velocities at short mean distances.}
	\end{center}
\end{figure}

The rotation velocities of the objects in a galaxy with pure Newtonian dynamics
is shown in Figure 4.
The rotation velocities in red and standard deviation with magenta are obtained in the time interval $t \in [0, 8\times 10^6]$.
The green curve is the function $v(r) = (680/r)^{0.5}$ for one object in circulation around a mass center with a mass M=680 times
the mass of the object.  The blue straight line is $v=0.15$ and it
corresponds to $v=220$ km/s for stars in the Milky Way \cite{Camarillo2018}.
The rotation velocities decline even more rapidly with the distance to the center of rotation than
given by $v(r)=(680/r)^{0.5}$, and in disagreement with the observed
mean rotation velocities of the stars in galaxies.  

The dynamics with MOGA increase the rotation velocities at distances $r> r_0$ to the center of the galaxy.
 The rotation velocities for MOGA dynamics and for different values of $r_0$ are shown in the next figure.
 The rotation velocities are for the bound objects with the distances $r < 100 000$.
The mean velocities of the objects in Figure 5 are determined in the time interval $t \in [0,8\times 10^6]$.
 They are constant within the accuracy of the simulations already for a modified attraction with $r_0=10000$ corresponding to
10 kpc for the Milky Way and a smaller value of the distance $r_0$ for the onset of the modification only increases the constant mean velocity.
So a modification of the ISL attraction to an inverse attraction not only stabilizes the galaxy, (Figure 3) but also increases the rotation
velocities at large distances and make them rather constant with respect to the distance to the center of rotation.

\begin{figure} 	
	\begin{center}	
\includegraphics[width=8cm,angle=-90]{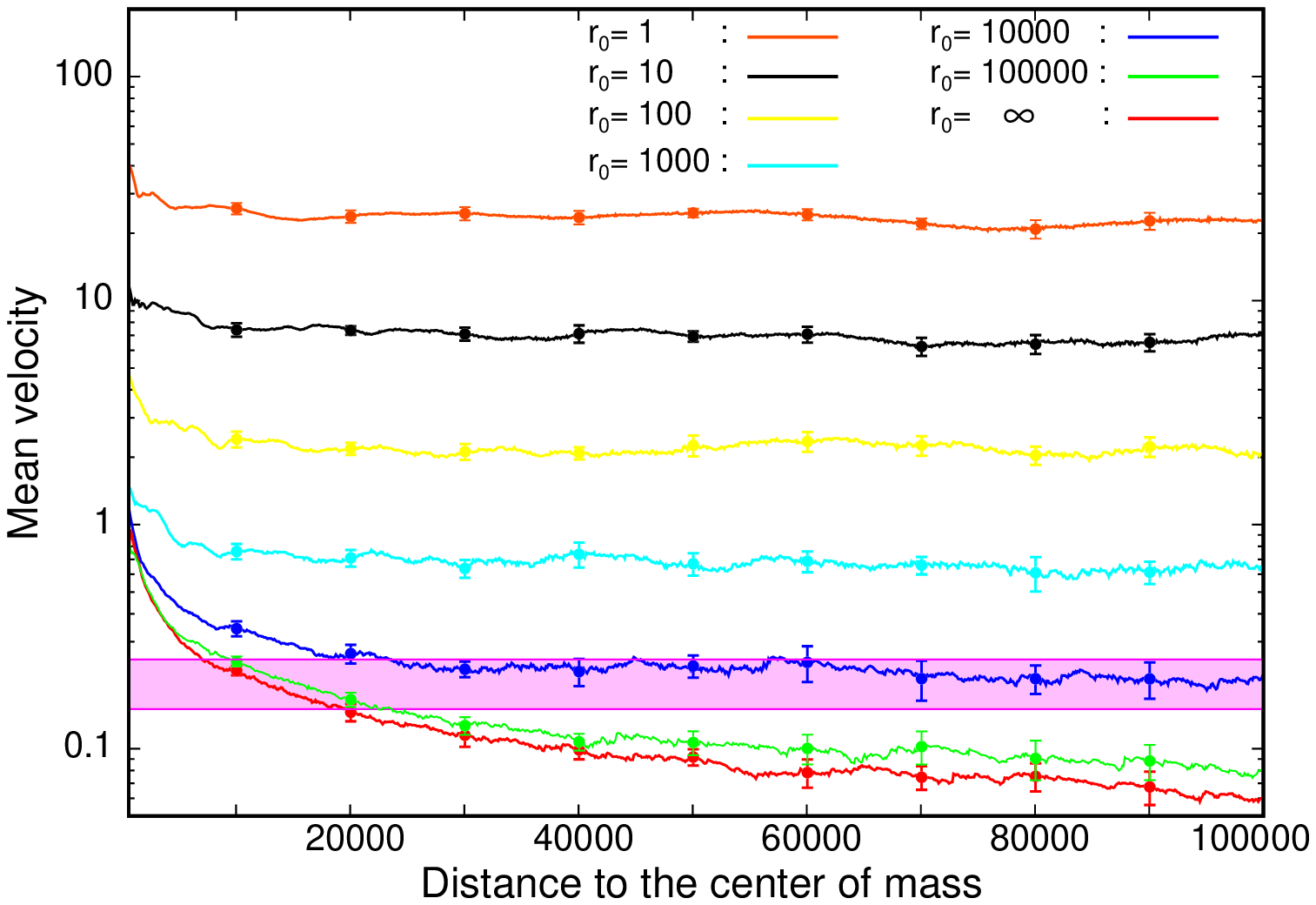}
		\caption{ The MOGA ($log-$)  rotation velocities of the objects in the galaxy models of the Milky Way 
as a function of the distances to their center of mass in the galaxies and for various values of the distance $r_0$
		for the onset of the inverse attraction. Distance is in parsec and the velocities are in MD units
		where a mean velocity $\bar{v} \approx 0.15-0.25$ corresponds $\bar{v}=220$ km s$^{-1}$ in the Milky Way.
		The velocity interval in magenta is for velocities that correspond to rotation
		velocities $v \in [0.15, 0.25]$.}
	\end{center}
\end{figure} 

The rotation velocity of stars in the Milky Way can be related to the rotation velocities in  the MOGA models of a galaxy.
The relation between lengths is obtained from the sizes of the galaxies, and the relation between velocities is obtained from the rotation time and velocity
in the Milky Way and the corresponding rotation time and velocities in the models of galaxies.
The rotation velocity of 
stars in the Milky Way at the distances $r \in [6.72, 8.40]$ kpc is $v(r)=220 \pm 10$  km s$^{-1}$ \cite{Camarillo2018}.
 An estimate of the mean velocity of the objects in the models for a galaxy is associated with uncertainty.
 The mean rotation velocity $\bar{v} \approx 0.15-0.25$ in the galaxy models is determined to  correspond to
 the mean  rotation velocity  $v=220$ km s$^{-1}$ 
 in the Milky Way \cite{Toxvaerd2022a},
 and the colored area in Figure 5 is for mean rotation velocities in the interval $v \in [0.15,0.25]$.
 The IA attractions 
   increase systematically the mean velocities of the baryonic objects,  and the  rotation velocities in the MOGA galaxies only agree with the
   observed  velocities in the Milky Way  for a large value of the modification distance $r_0$.
 The simulations indicate that the stability of a galaxy with a rather constant rotation velocity which corresponds to $\approx$ the rotation velocity in the Milky Way can be obtained by a modification of the attractions
 for far away objects with distances corresponding to $r \ge 10$ kpc (blue curve in Figure 5). The estimation is with a big uncertainty, but
 one can, however, exclude a significantly shorter value of the modification range $r_0$. A smaller value than $r_0$= 10 kpc 
 will according to the results in Figure 5 lead to an unrealistic
 high rotation velocity.
 The  enhanced accelerations (MOND) or attractions (MOGA) for $r_0 \le 1000 \rightarrow a_0 \ge 1.0\times 10^{-6}$ result in a too high mean rotation velocity for MOGA, and in the case of MOND, the galaxy is spontaneously 
 destabilized.

\section{Conclusion}
\subsection{ Summary of the simulations}

The Universe is more than thirteen billion years old and the  Milky Way was created relatively shortly after the birth of the Universe.
 An appropriate time unit for the evolution of the Universe is, however, not a year but the time  of one rotation of a galaxy.
 The Milky Way has rotated about sixty times with  $\approx$ 240 million years per rotation \cite{Camarillo2018}, and
 a Newtonian system of stars is rather stable even after the double number of rotations \cite{Toxvaerd2022a}. But the 
 rotation velocity of the objects in the  Newtonian system with  ISL attraction is not constant but declines proportional to the distance $r$ from
 the center of mass as $r^{-0.5}$ (Figure 4) and inconsistent with the rate of rotation of galaxies \cite{Corbelli2000,Gentile2011,Camarillo2018}. 

MOGA with the modification of the gravitational attraction from a Newtonian ISL attraction to an IA attraction for pairs of baryonic objects at
distances $r \ge r_0$  stabilizes the objects in the galaxy further (Figure 2 and Figure 3), and the modification changes the mean rotation
velocity of objects in the galaxy from a classical Newtonian/Kepler behavior to a rather constant velocity to the distance to the center of rotation
in the galaxy (Figure 5). The rotation velocity in the galaxies with MOGA is in qualitative agreement with the observed rotation velocity in the
Milky Way and galaxies in the Universe.
\subsection{Discussion}

Newton began $Principia$ by postulating his three laws, and then he derived the second law, Eq. (1), from a discrete algorithm which today
is the most used algorithm in computer simulations. The algorithm is derived from the $action$ of a \textit{force quant} \cite{Toxvaerd2022}, more than three hundred years before the
formulation of Quantum Mechanics, Feynman's path integral formulation,  the General Relativity theory (GR), and quantum electrodynamics (QED).
Newton's formulation of the ISL of universal gravitation, Eq (2), was derived much later in $Principia$ , on page 401 in the Third Book \textit{Mundi Systemate}, and
where Newton first formulated four philosophic principles for Natural Science. The first and in an English translation reads:\\
\textit{ We are to admit no more causes of natural things than such as are both true and sufficient to explain their appearances.}\\
 MOGA acknowledges this principle. But MOND does not obey Newton's third law, it does not conserve momentum and
angular momentum (Figure 6) and the MOND galaxies are unstable (Figure 3).

The present results raise some questions: Can the dynamics of the Universe be explained alone by a modification of Newton's laws, his universal gravitation, and QED, or
do one also need explicitly to include an effect of dark matter in the dynamics?\\
And if a modification of the ISL law is sufficient, what causes this modification and how 
should this modification be formulated?\\
None of the two questions can of cause be answered definitively from the present investigation.

 Modifications of the Newtonian ISL  to an IA-like attraction have been proposed for a long time 
 in an attempt to obtain the stability of galaxies
  by mean of the standard model \cite{Fischbach2001,Adelberger2003,Henrichs2021,Capozziello2017,Finch2018}. The modifications are typically obtained by a Yukawa-force correction, Eq. (13).
Milgrom's interpolation formulae for a single pair of baryonic objects leads to an asymptotic modification of the forces, given by Eq. (11). An example of the two modifications
are shown in Figure 1 for specific values of the constants for the modifications.
Another attempt to modify the dynamics of galaxies is the \textit{f(R)} \cite{Buchdahl1959}, \textit{ f(T)} \cite{Einstein1928} ,
and \textit{f(Q)}  \cite{Jimenez2020}  theories
where the modified behaviour is obtained by modifying Einstein's  GR theory. For a review of the  \textit{f(R)}  modifications see the reviews\cite{Capozziello2011,Capozziello2024}.
Here I would like to propose another possibility, viz
    gravitational lensing \cite{Bartelmann2010,Abbott2016,Mukherjee2021} and focusing, caused by heavy centers of mass in the galaxy
     of the gravitational waves from objects in the galaxy that are located at long distances from the object.

\appendix
\section{The discrete algorithm} 
The  classical mechanical simulations of the dynamics of interacting objects are performed by 
using Newton's  algorithm for discrete classical dynamics \cite{Newton1687}.
Simulations of molecular and atomic systems are named ''Molecular Dynamics'' (MD), and
 almost all  MD simulations  and many simulations in celestial mechanics are performed using Newton's algorithm, but  
 with the name ''Leap-frog'' or  the ''Verlet algorithm''  \cite{Verlet1967}, and the 
 algorithm also appears under a variety of other names.  It was, however, 
  Isaac Newton who first formulated the \textit{Discrete Molecular Dynamics} algorithm, when he in
  PHILOSOPHI\AE \ NATURALIS PRINCIPIA MATHEMATICA $(Principia)$  derived his second law for classical mechanics
  \cite{Newton1687,Toxvaerd2020,Toxvaerd2023}.

\subsection{Newton's Discrete Molecular Dynamics algorithm}

  In Newton's  discrete dynamics  a new  position $\textbf{r}_i(t+\delta t)$ at time $t+\delta t$ of an object
$i$ with the mass $m_i$  is determined by
the force $\textbf{f}_i(t)$ acting on the object   at the discrete positions $\textbf{r}_i(t)$  at time $t$, and  
 the position $\textbf{r}_i(t-\delta t)$ at $t - \delta t$  as
\begin{equation}
	 m_i\frac{\textbf{r}_i(t+\delta t)-\textbf{r}_i(t)}{\delta t}
			=m_i\frac{\textbf{r}_i(t)-\textbf{r}_i(t-\delta t)}{\delta t} +\delta t \textbf{f}_i(t),	
 \end{equation}
where the momenta $ \textbf{p}_i(t+\delta t/2) =  m_i (\textbf{r}_i(t+\delta t)-\textbf{r}_i(t))/\delta t$ and
 $  \textbf{p}_i(t-\delta t/2)=  m_i(\textbf{r}_i(t)-\textbf{r}_i(t-\delta t))/\delta t$ are constant in
the time intervals in between the discrete positions.
Newton  begins \textit{Principia} by postulating Eq. (A.1) in \textit{Proposition I},
and he obtained his second   law   as the  limit $ lim_{\delta t \rightarrow 0}$ of the equation.

 Usually, the algorithm, Eq. (A.1), is  presented  as the Leap-frog algorithm for the velocities
\begin{equation}
\textbf{v}_i(t+\delta t/2)=  \textbf{v}_i(t-\delta t/2)+ \delta t/m_i  \textbf{f}_i(t),
\end{equation}
 and the positions
are determined from the discrete values of the momenta/velocities as
\begin{equation}
\textbf{r}_i(t+\delta t)= \textbf{r}_i(t)+ \delta t \textbf{v}_i(t+\delta t/2).	  
\end{equation}	  

The rearrangement of Eq. (A.1) gives  the Verlet algorithm \cite{Verlet1967}

\begin{equation}
	\textbf{r}_i(t+\delta t)=2\textbf{r}_i(t)-\textbf{r}_i(t-\delta t) +\delta t \textbf{f}_i(t)^2/m_i .	  
\end{equation}

\subsection{The invariances in Classical Mechanics, Discrete  Molecular Dynamics, MOND and MOGA}

Classical analytic dynamics are time-reversible and symplectic and a conservative system of $N$ baryonic objects has
three invariances: conserved momentum, angular momentum, and energy. Newton's discrete Molecular Dynamics is
also reversible and symplectic and has the same invariances. MOGA and MOND are time-reversible and symplectic,
but only MOGA maintains the three invariances, whereas MOND does not ensure momentum and angular momentum conservation. 
The proof is given below.

 Newton's discrete dynamics for a  system of  $N$ spherically symmetrical objects
     with masses $ m^N \equiv m_1, m_2,..,m_i,..,m_N$ and positions   \textbf{r}$^N(t) \equiv$ \textbf{r}$_1(t)$, \textbf{r}$_2(t)
     ,..,$\textbf{r}$_i(t),..$\textbf{r}$_N(t)$  is obtained 
 by Eqn. (A.1). Let the force, $ \textbf{F}_i$ on object No $i$ be a sum of pairwise  forces  $ \textbf{f}_{ij}$ between pairs of   objects $i$ and $j$
 \begin{equation}
	 \textbf{F}_i=  \sum_{j \neq i}^{N} \textbf{f}_{ij}.
 \end{equation}	

   Newton's discrete dynamics, Eq. (A.1) is a central difference algorithm and it is time symmetrical, so the discrete dynamics is time reversible and symplectic \cite{Friedman1991}. MOND and MOGA with Eq. (A.1) are also time reversible and symplectic.
  
 The momentum for a conservative system with the discrete dynamics, Eq. (A.1) of the $N$ objects is conserved since 
\begin{eqnarray}
	\sum_i^N m_i\frac{\textbf{r}_i(t+\delta t/2)-\textbf{r}_i(t)}{\delta t}= 
\sum_i^N \textbf{p}_i(t+\delta t/2)= \\ \nonumber
\sum_i^N \textbf{p}_i(t-\delta t/2)+ \delta t\sum_{i,j \neq i}^N \textbf{f}_{ij}(t)=
\sum_i^N \textbf{p}_i(t-\delta t/2),
\end{eqnarray}
where $\sum_{i,j \neq i}^N \textbf{f}_{ij}(t)=0$ with $ \textbf{f}_{ij}(t)=-\textbf{f}_{ji}(t)$ due to Newton's third law. But only the discrete dynamics and MOGA conserve the
 momentum, whereas MOND does not  because (see Eq. (8))  \cite{Felten1984}
 \begin{equation} 
	 \textbf{f}_{ij}(t)\sqrt{1+4 m_i a_0/\mid F_i\mid} \neq   -\textbf{f}_{ji}(t)\sqrt{1+4 m_j a_0/\mid F_j\mid}.
 \end{equation}	 
The shortcoming of MOND with respect to momentum conservation is independent of the algorithm because 
the momentum conservation in analytic dynamics is also ensured by
  $\sum_{i,j \neq i}^N \textbf{f}_{ij}(t)=0$. 

The discrete positions and momenta are not
 known simultaneously. An expression for the angular momentum of the conservative system is 
\begin{eqnarray}
	\textbf{L}(t)=\sum_i^N \textbf{r}_i(t)  \times (\textbf{p}_i(t+\delta t/2) +\textbf{p}_i(t-\delta t/2) )/2 \nonumber \\
	= \sum_i^N \textbf{r}_i(t)  \times  (m_i \textbf{r}_i(t+\delta t) -m_i \textbf{r}_i(t-\delta t) )/ 2\delta t   ).
\end{eqnarray}	
The angular momentum is conserved since (using $\textbf{r}_i(t)  \times ( \textbf{f}_{ij}(t)+\textbf{f}_{ji}(t)) =0,$
  $\textbf{a} \times \textbf{a}=0,$ $ \textbf{a} \times \textbf{b}=-  \textbf{b} \times \textbf{a}$ and  Eq. (A.4)  )
\begin{eqnarray} 
 2 \delta t \textbf{L}(t)=\sum_i^N \textbf{r}_i(t) \times (m_i \textbf{r}_i(t+\delta t) -m_i \textbf{r}_i(t-\delta t) )  \nonumber \\
=\sum_i^N m_i \textbf{r}_i(t) \times (2\textbf{r}_i(t) -2\textbf{r}_i(t-\delta t) )  
=\sum_i^N m_i\textbf{r}_i(t-\delta t) \times (\textbf{r}_i(t) +\textbf{r}_i(t))  \nonumber \\
=\sum_i^N m_i\textbf{r}_i(t-\delta t) \times (\textbf{r}_i(t) -\textbf{r}_i(t-2\delta t))
=2 \delta t \textbf{L}(t-\delta t).
\end{eqnarray}	
MOGA fulfils Newton's third law with $\sum_{i,j \neq i}^N \textbf{f}_{ij}(t)=0$  and conserves the angular moment whereas MOND does not conserve this invariance.

 The  energy in analytic dynamics is the sum of potential energy $ U(\textbf{r}^N(t))$ and kinetic energy $K(t)$, 
 and it is an invariance for a conservative system.
 The kinetic energy at time $t$ in the discrete dynamics is, however, ill-defined since the velocities change at time $t$.
The energy invariance
in the discrete dynamics can, however,  be seen by considering the change in kinetic energy, $\delta K([t-\delta t/2, t+\delta t/2])$ and
potential energy and  $\delta U ([t-\delta t/2, t+\delta t/2])$
in the time  interval $[t-\delta t/2, t+\delta t/2]$.

The loss in  potential energy, $-\delta U$ is defined as
the work done by the forces at a  move of the positions \cite{Goldstein}. 
An expression for the work, $W$ done in the time interval by the discrete dynamics
from the position  $(\textbf{r}_i(t)+ (\textbf{r}_i(t-\delta t))/2$ at $t-\delta t/2$
to the position  $(\textbf{r}_i(t+\delta t)+ \textbf{r}_i(t))/2$ at $t+\delta t/2$ 
is  \cite{Toxvaerd2023}
\begin{eqnarray}
	-\delta U= W= \sum_i^N  \textbf{f}_i(t)(\frac{\textbf{r}_i(t+\delta t)+ \textbf{r}_i(t)}{2}
	-\frac{\textbf{r}_i(t)+ \textbf{r}_i(t-\delta t)}{2})\\ \nonumber
	= -\sum_i^N  \textbf{f}_i(t)(\textbf{r}_i(t+\delta t) -\textbf{r}_i(t-\delta t))/2.
\end{eqnarray}	
By rewriting Eq. (A.4) to
\begin{equation}
	\textbf{r}_i(t+ \delta t) -\textbf{r}_i(t-\delta t)= 2(\textbf{r}_i(t) -\textbf{r}_i(t-\delta t))+\frac{\delta t^2}{m_i} \textbf{f}_i(t),
\end{equation}
and inserting in Eq. (A.10) one obtains an expression for the total work in the time interval
\begin{equation}
	-\delta U=  W=  \sum_i^N  [(\textbf{r}_i(t) -\textbf{r}_i(t-\delta t)) \textbf{f}_i(t)+ \frac{\delta t^2}{2m_i}\textbf{f}_i(t)^2].
\end{equation}

The change in kinetic energy in the time interval $[t-\delta t/2, t+\delta t/2]$ is
\begin{eqnarray}
	\delta K= \sum_i^N \frac{1}{2}m_i [\frac{(\frac{\textbf{r}_i(t+\delta t)+\textbf{r}_i(t)}{2}-\textbf{r}_i(t))^2}{(\delta t/2)^2}-
	\frac{(\textbf{r}_i(t) -\frac{\textbf{r}_i(t) + \textbf{r}_i(t-\delta t)}{2})^2}{(\delta t/2)^2}] \\ \nonumber
	= \sum_i^N \frac{1}{2}m_i [\frac{(\textbf{r}_i(t+\delta t) -\textbf{r}_i(t))^2}{\delta t^2}-
\frac{(\textbf{r}_i(t) -\textbf{r}_i(t-\delta t))^2}{\delta t^2}].
\end{eqnarray}
By rewriting  Eq. (A.4) to
\begin{equation}
	\textbf{r}_i(t+ \delta t) -\textbf{r}_i(t)= \textbf{r}_i(t) -\textbf{r}_i(t-\delta t)+\frac{\delta t^2}{m_i} \textbf{f}_i(t)
\end{equation}
   and inserting the squared expression for  $\textbf{r}_i(t+\delta t) -\textbf{r}_i(t)$ in  Eq. (A.13), the change in kinetic energy
   is
\begin{equation}
	\delta K= \sum_i^N [ (\textbf{r}_i(t)-\textbf{r}_i(t - \delta t))\textbf{f}_i(t) +\frac{\delta t^2}{2m_i} \textbf{f}_i(t)^2].
\end{equation}
The energy invariance in Newton's discrete dynamics is expressed by Eqn. (A.12),  and  Eq. (A.15) as \cite{Toxvaerd2023}
\begin{equation}
	\delta E=\delta U+\delta K=0.
\end{equation}
The energy invariance is due to the time symmetry,  and it is valid for any discrete force.
It does not rely on the existence of an analytic force with an analytic potential.

 \begin{figure} 	
	\begin{center}	
\includegraphics[width=8cm,angle=-90]{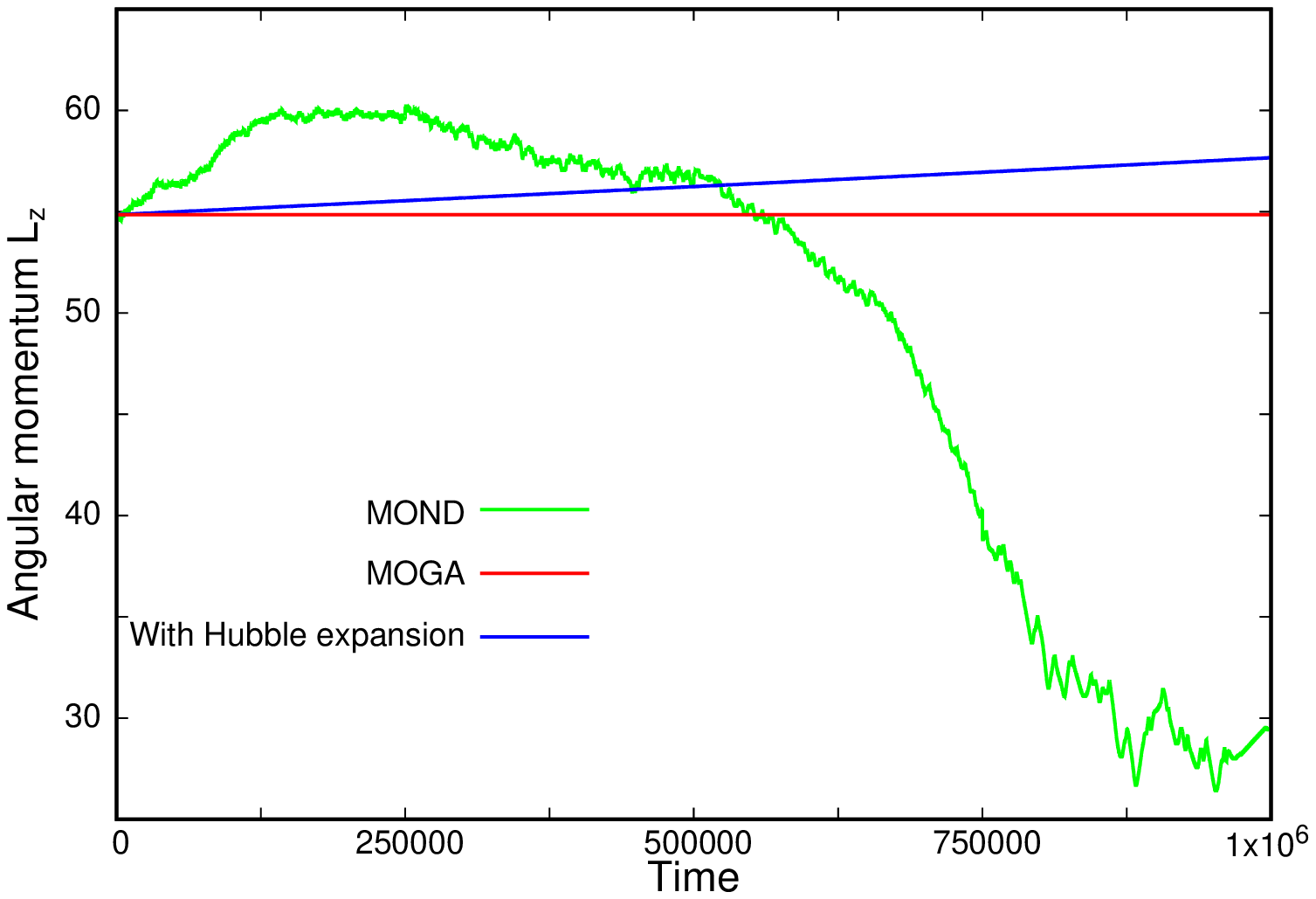}
	\caption{ Time evolution of the z-component, $\textrm{L}_z$,  of the angular momentum of the galaxies. In green is $\textrm{L}_z$
		for MOND with $a_0=10^{-10}$. The red line is the conserved  $\textrm{L}_z$ for
		the Newtonian and MOGA  dynamics without a Hubble expansion. The corresponding evolution of   $\textrm{L}_z$  with
		the Hubble expansion included is shown in blue.}
	\end{center}
\end{figure}

\subsection{ The simulations with Discrete  Molecular Dynamics, MOND and MOGA}

The simulations are started with various start configurations of positions $\textbf{r}(0)^N$  and with the small time increment
$\delta t=0.0025$. Each simulations  are performed for $3.2 \times 10^9$ time steps corresponding to a reduced time
$t=8.0 \times 10^6 \approx$ 13.4 Gyr \cite{Toxvaerd2022}. The exact algorithm is absolutely stable and all the simulations
are performed without any constraints or adjustments. The algorithm and the explanation of the stability of Newton's discrete algorithm
are explained in a recent review of Discrete Molecular Dynamics \cite{Toxvaerd2023}.

MOND is simulated with the modified acceleration, Eq. (8), and MOGA with the modified forces, Eq.(11).
All the simulations conserve energy,  Newtonian dynamics, and MOGA  conserve also momentum and angular momentum, but
MOND does not.
Figure A1 shows the time evolution for $4. \times 10^8$ time-steps in  green  of the
z-component of the angular momentum with  MOND dynamics for the galaxies shown in the previous figures, together with the corresponding 
components  with pure Newtonian or MOGA dynamics and without Hubble expansion in red, and with the Hubble
expansion in blue. The simulations are started with a disk-like configuration $\textbf{r}(0)^N$  
with  $\textrm{L}_x \approx 0$ and  $\textrm{L}_y \approx 0$
and with  $\textrm{L}_z=54.85$. 
The simulation with MOND in green is for the  small value $a_0= 10^{-10}$
with the number of objects in the galaxy shown in green in Figure 3. 
The momentum and  the angular momentum with MOND vary with time, whereas
the momentum and  angular momentum with  Newtonian dynamics and with MOGA without
the Hubble expansion is conserved. The Hubble expansion and with Newtonian dynamics or with MOGA (blue line)
increases the angular momentum monotonically, but very slowly. The MOND  angular momentum shown in Figure A1 is
for $a_0= 10^{-10}$. 
The values of the angular momentum for larger values of $a_0$ fluctuate with amplitudes
that are decades bigger and the objects are released from  the galaxy (Figure 3).\\
\\
$   $ \\
$\textbf{Acknowledgements}$
This work was supported by the VILLUM Foundation Matter project, grant No. 16515.
\\
$   $ \\
$\textbf{Data Availability Statement}$ Data will be available on request.
\\


\begin{thebibliography}{99}
	\bibitem{Swart2017}  J. G. de Swart, G. Bertone and J.van Dongen,    Nat. Astron.,  {\bf 1} 0059 (2017).
	\bibitem{Bullock2017} J. S. Bullock and M. Boylan-Kolchin, Annu. Rev. Astron. Astrophys.,   {\bf 55} 343 (2017).
	\bibitem{Perivolaropoulos2022} L. Perivolaropoulos  and F. Skara,  New Astron. Rev.,    {\bf 95} 101659 (2022).
	\bibitem{Fischbach2001} E. Fischbach, D. E. Krause, V. M. Mostepanenko  and M. Novello, Phys. Rev. D,   {\bf 64}  075010 (2001).
	\bibitem{Adelberger2003} E. G. Adelberger, B. R. Heckel and A. E. Nelson, Annu. Re. Nucl. Part. Sci.,   {\bf 53} 77 (2003). 
	\bibitem{Lee2020} J. G. Lee,  E. G. Adelberger, T. S. Cook, S. M. Fleischer  and  B. R. Heckel, Phys. Rev. Lett.,   {\bf 124}  101101 (2020).
	\bibitem{Henrichs2021} J. Henrichs, M. Lembo, F. Iocco  and L. Amendola, Phys. Rev. D,   {\bf 104} 043009 (2021).
	\bibitem{Rubin1980} V. C. Rubin,  W. K. Ford Jr. and N. Thonnard, Astrophys. J., {\bf 238} 471 (1980).
	\bibitem{Rubin1985} V. C. Rubin, D. Burstein,  W. K. Ford Jr. and N. Thonnard N 1985 {\it Astrophys. J. } {\bf 289} 81 (1985).
	\bibitem{Gentile2011} G .Gentile, B. Famaey and  W. J. G. de Blok, A\&A,  {\bf 527} A76 (2011). 
	\bibitem{Corbelli2000} E. Corbelli  and P. Salucci, Mon. Not. R. Astron. Soc., {\bf 311} 441 (2000).
	\bibitem{Bertone2018} G. Bertone, D. Hooper, Rev. Mod. Phys., {\bf 90} 045002 (2018).
	\bibitem{Milgrom1983} M. Milgrom, Apj, {\bf 270} 371 (1983). 
	\bibitem{Bekenstein1984} J. D. Berkenstein  and M. Milgrom, Astrophys. J., {\bf 286} 7 (1984).
	\bibitem{Famaey2012} B. Famaey and  S. McGaugh, Living Rev. Relativity, {\bf 15} 10 (2012).
	\bibitem{Capozziello2011}  S. Capozziello and  M. De Laurentis,  Phys. Rep.  {\bf 509} 167 (2011).
	\bibitem{Clifton2012} T. Clifton,  P. G. Ferreira, A. Padilla and C.Skordis, Phys. Rep., {\bf 513} 1 (2012). 
	\bibitem{Mendoza2015} S. Mendoza, Can. J. Phys., {\bf 93} 217 (2015).
	\bibitem{Bartelmann2010} M. Bartelmann, Clas. Quantum Grav., {\bf 27} 233001 (2010).
	\bibitem{Mukherjee2021} S.Mukherjee,  B. D. Wandelt, S. M. Nissanke  and A. Silvestri, Phys. Rev. D, {\bf 103} 043520 (2021).
	\bibitem{Abbott2016} B. P. Abbott et al., Phys. Rev. Lett., {\bf 116} 061102 (2016).
	\bibitem{Boran2018} S. Boran, S. Desai, E. O. Kahya and R. P. Woodard, Phys. Rev. D, {\bf 97} 041501(R) (2018).	
	\bibitem{Hockney1974} R. W. Hockney, S. P. Goel and  J. W. Eastwood, J. Comput. Phys., {\bf 14} 148 (1974). 
	\bibitem{Klypin1983} A. A. Klypin and  S. F. Shandarin, MNRAS, {\bf 204} 891 (1983).
	\bibitem{Centrella1983} J. Centrella  and A. L. Melott, Nature,  {\bf 305} 196 (1983).
	\bibitem{Zeldovich1970} Ya. B. Zeldovich, Astrom. \& Astrophys., {\bf 5} 84 (1970).
	\bibitem{Springel2005} V. Springel, MNRAS, {\bf 364} 1105 (2005). 
	\bibitem{Price2018} D. J. Price et al, Publ. Astron. Soc. Aust., {\bf 35} e031 (2018). 
	\bibitem{Weinberger2020} R. Weinberger, V.Springel  and R. Pakmor, Astrophys. J., Suppl. Ser., {\bf 248} :32 (2020).
	\bibitem{Schaye2010} J. Schaye  et al., MNRAS, {\bf 402} 1536 (2010). 	
	\bibitem{Dubois2014} Y. Dubois, M.Volonteri  and J. Silk, MNRAS, {\bf 440} 1590 (2014). 
	\bibitem{Vogelsberger2014} M. Vogelsberger  et al., MNRAS, {\bf 444} 1518 (2014).
	\bibitem{Dubois2016} Y. Dubois  et al., MNRAS, {\bf 463} 3948 (2016). 
	\bibitem{Ludlow2021} A. D. Ludlow, S. M. Fall, J.Schaye  and D. Obreschkow, MNRAS, {\bf 508} 5114 (2021). 
	\bibitem{Angus2014} G. W. Angus, G. Gentile, A. Diaferio, B. Famaey and K.J. van der Heyden,  MNRAS, {\bf 440} 746 (2014).
	\bibitem{Angus2014a} G. W. Angus, A. Diaferio, B. Famaey and K.J. van der Heyden, J. Astrophys. Astron., {\bf 10} 079 (2014).
	\bibitem{Lughausen2014} F. L\"{u}ghausen, B. Famaey and P Kroupa, MNRAS, {\bf 441} 2497 (2014). 
	\bibitem{Vogelsberger2020} M. Vogelsberger, F. Marinacci, P.Torrey and E. Puchwein, Nat. Rev. Phys, {\bf 2} 42 (2020).
	\bibitem{Newton1687} I. Newton , PHILOSOPHI\AE \ NATURALIS PRINCIPIA MATHEMATICA. \textit{LONDINI}, \textit{Anno} MDCLXXXVII.  Second Ed.1713; Third Ed. (1726)	
	\bibitem{Toxvaerd2020} S. Toxvaerd, Eur. Phys. J. Plus, {\bf 135} 267 (2020).
	\bibitem{Toxvaerd2022} S. Toxvaerd, Eur. Phys. J. Plus, {\bf 137} :99 (2022). 
	\bibitem{Toxvaerd2022a} S. Toxvaerd, Class. Quantum Grav., {\bf 29} 22500 (2022).
	\bibitem{Toxvaerd2023} S. Toxvaerd, Comprehensive Computational Chemistry, {\bf 3} 329 (2023). 
	\bibitem{Felten1984} J. E. Felten, Astron. J., {\bf 286} 3 (1984).	
        \bibitem{Capozziello2017}  S. Capozziello, P. Jovanovi\'{c}, V. B. Jovanovi\'{c} and D. Borka, J. Cosmol. Astropart. P. {\bf 06} 044 (2017).
	\bibitem{Finch2018} A. Finch and J. L. Said, Eur. Phys. J. C  {\bf 78} 560 (2018).
	\bibitem{Benistry2023} D. Benistry, A.-C. Davis and N. W. Evans, Astrophys. J. Lett. {\bf 953} L2 (2023).
	\bibitem{Benistry2023a} D. Benistry and S. Capozziello, Phys. Dark Univ.  {\bf 39} 101175 (2023).
        \bibitem{Capozziello2024}  S. Capozziello, M. Capriolo and S. Nojiri,  Phys. Lett. B  {\bf 850} 138510 (2024).
        \bibitem{Cardone2011} V. F. Cardone and  S. Capozziello, MNRAS, {\bf 414} 1301 (2011).
	\bibitem{Chen2016} Y-J. Chen, W. K. Tham, D. E. Krause, D. L\'{o}pez, E. Fischback and R. S. Decca, Phys. Rev. Lett., {\bf 116} 221102 (2016). 
	\bibitem{Bimonte2021} G. Bimonte, B. Spreng, P. A. Maia Neto, G-L. Ingold, G. L. Klimchitskaya, V. M. Mostepanenko  and R. S. Decca, Universe, {\bf 7} 93 (2021).
	\bibitem{Baeza-Ballesteros2022} J. Baeza-Ballesteros, A. Donini  and  S. Nadal-Gisbert, Eur. Phys. J. C, {\bf 82}: 154 (2022). 	
	\bibitem{Rix2013} H-W. Rix  and J. Bovy, Astron. Astrophys. Rev., {\bf 21} 61 (2013). 	
	\bibitem{Deason2020} A. J. Daeson  et al., MNRAS, {\bf 496} 3929  (2020).
	\bibitem{Li2021} Z-Z. Li  and J. Han, Astrophys. J. Lett., {\bf 915} L18 (2021). 
	\bibitem{Soltis2021} J. Soltis, S. Casertano  and A. G. Riess, Astrophys. J. Lett.,  {\bf 908} L5 (2021). 	
        \bibitem{Jiao2023} Y. Jiao, F. Hammer, H. Wang, J. Wang, P. Amram, L. Chemin and Y. Yang, A\&A,  {\bf 678} A208 (2023).
	\bibitem{Aerseth1963} S. J. Aerseth, MNRAS, {\bf 126} 223 (1963).
	\bibitem{Gupta2012} A. Gupta, S. Mathur, Y. Krongold, F. Nicastro  and M. Galeazzi, Astrophys. J. Lett., {\bf 756} :L8 (2012).
	\bibitem{Bergman2018} J. N. Bergman, M. E. Anderson, M. J. Miller, E. Hodges-Kluck, X. Dai, J-T. Li, Y. Li and Z. Qu, Astrophys. J., {\bf 862} :3 (2018).
	\bibitem{Famaey2005} B. Famaey  and J. Binney, MNRAS, {\bf 363} 603	(2005).
	\bibitem{Camarillo2018} T. Camarillo, P. Dredger  and B. Ratra, Astrophys. Space Sci., {\bf 363} :268 (2018). 
        \bibitem{Buchdahl1959} H. A. Buchdahl, Phys. Rev.,  {\bf 116}, 1027 (1959). 
        \bibitem{Einstein1928} A. Einstein, Sitzungsber. Preuss. Akad. Wiss. Phys. Math. Kl. 217 (1928); 401 (1930); Math. Ann. 102, 685 (1930).
	\bibitem{Jimenez2020} J. B. Jim\'{e}nez, L. Heisenberg, T. Koivisto and S. Pekar, Phys. Rev. D,  {\bf 101}, 103507 (2020. 
	\bibitem{Verlet1967} L. Verlet, Phys. Rev.,  {\bf 159}, 98 (1967). 
	\bibitem{Friedman1991}  A. Friedman and S. P. J. Auerbach, J. Comput. Phys.,  {\bf 93} 177, $ibid$  {\bf 93} 189 (1991).
	\bibitem{Goldstein} H. Goldstein,  {\it Classical Mechanics}, (Addison-Wesley Press Second Ed. 1980), Chap. 1.
	\end{thebibliography}
\end{document}